\documentclass[12pt]{article}
\usepackage{epsf}
\usepackage{graphicx}

\renewcommand{\thefootnote}{\#\arabic{footnote}}

\newcommand{\gtrsim}{ \mathop{}_{\textstyle \sim}^{\textstyle >} }

\begin{document}

\setcounter{footnote}{0}
\begin{titlepage}

\begin{center}

\vskip .5in

{\Large \bf
Cosmological Constraints on Isocurvature and Tensor Perturbations
}

\vskip .45in

{\large
 Masahiro Kawasaki 
 and Toyokazu Sekiguchi
  \\
}

\vskip .45in

{\em
Institute for Cosmic Ray Research,
University of Tokyo\\
Kashiwa 277-8582, Japan, \\
}

\end{center}

\vskip 1in

\begin{abstract}
We investigate cosmological constraints on primordial isocurvature and
tensor perturbations, using recent observations of the cosmic microwave
background and the large scale structure. We find that present
observations are consistent with purely adiabatic initial conditions for
the structure formation under any priors on correlations of isocurvature
modes, and upper limits on the contribution of isocurvature and tensor
perturbations are presented.  We also apply the obtained constraints to
some specific theoretical models, axion isocurvature perturbation models
and curvaton models, and give some implications for theoretical models.
\end{abstract}

\end{titlepage}

\renewcommand{\thepage}{\arabic{page}}
\setcounter{page}{1}
\renewcommand{\thefootnote}{\#\arabic{footnote}}
\renewcommand{\theequation}{\thesection.\arabic{equation}}

\section{Introduction}\label{sec:introduction}

\setcounter{equation}{0}

Recent cosmological observations, such as the cosmic microwave
background (CMB) and the large scale structure (LSS) provide us
information on primordial perturbations which seed the structure of the
present universe.  All observations suggest that the primordial
fluctuation is almost adiabatic and scale-invariant
\cite{Spergel:2006hy,Tegmark:2006az}.  Inflation is the most promising
mechanism to generate the scale-invariant adiabatic fluctuation in the
early universe.  On the other hand, primordial isocurvature
perturbations are also generally generated, along with the tensor
perturbations, in the inflation universe. Many possible sources and
mechanisms generating isocurvature perturbations are known such as
axion, curvaton scenarios \cite{Lyth:2001nq,Moroi:2001ct} and
multi-field inflation models.
 
Therefore it is expected that constraints on primordial isocurvature and
tensor perturbations give us some useful information to build realistic
inflation models and models in particle physics.  Thus, constraints on
primordial isocurvature perturbations have been investigated by many
authors
\cite{Pierpaoli:1999zj,Enqvist:2000hp,Trotta:2001yw,Trotta:2002iz,
Valiviita:2003ty,Crotty:2003rz,Moodley:2004nz,Beltran:2004uv} (For
recent constraints we refer to
\cite{Bean:2006qz,Trotta:2006ww,Keskitalo:2006qv}).  However, there have
been few investigations on cosmological models with both isocurvature
and tensor perturbations.  This is partly because in most inflation
models tensor perturbations are expected to be small when (especially
correlated) isocurvature perturbations are generated.  However still
some models predicts generation of both isocurvature and tensor
perturbations \cite{Bartolo:2001rt,Byrnes:2006fr}.  From
phenomenological point of view, it is worth checking whether
cosmological observations are consistent with purely adiabatic initial
conditions even if we consider both isocurvature and tensor
perturbations.

In this paper we investigate constraints on cosmological models with
both isocurvature and tensor perturbations in light of cosmological
observations of CMB and LSS. We use data from two recent cosmological
observations, CMB temperature and polarization power spectra from WMAP
3-year result (WMAP3) and galaxy power spectrum from SDSS data release 4
of luminous red galaxy sample (SDSS DR4 LRG).  We only consider models
with one isocurvature mode along with adiabatic and tensor modes, which
are simple but suggestive for various models predicting generation of
isocurvature and tensor modes.  We investigate the isocurvature mode by
using three different priors on correlation between isocurvature and
adiabatic modes; 1) uncorrelated, 2) totally correlated and 3) generally
correlated models.  The reason why we investigate uncorrelated and
totally correlated models separately is that there are some simple
models predicting definite correlations. For examples, the axion
isocurvature perturbation model produces uncorrelated isocurvature mode
and the totally correlated one is predicted in curvaton scenarios.

The structure of the paper is as follows.  In
Section~\ref{sec:initial_condition} we briefly review the general
initial perturbations of the structure formation which includes the
isocurvature and tensor perturbations and we also gives the
parametrization used to constrain the isocurvature perturbations there.
In section~\ref{sec:models} we give some examples of models with both
isocurvature and tensor perturbations which are based on inflation
scenarios.  In section~\ref{sec:method} we show the method to obtain the
constraints on the isocurvature and tensor perturbations from the
combined set of cosmological observations.  In
section~\ref{sec:constraints} we present constraints on isocurvature and
tensor perturbations from CMB and LSS.  In section~\ref{sec:application}
we apply the obtained constraints on the isocurvature perturbation to
some specific models; axion isocurvature perturbation models and curvaton
scenarios.  Section~\ref{sec:conclusion} is dedicated to conclusions and
discussions.

\section{Initial perturbations for the structure formation}
\label{sec:initial_condition}
Scalar perturbations are generally decomposed into five
modes~\cite{Bucher:1999re}; adiabatic mode (AD), CDM isocurvature mode
(CI), baryon isocurvature mode (BI), neutrino isocurvature density
mode (NID) and neutrino isocurvature velocity mode (NIV). In the
framework of the linear perturbation theory, each mode evolves
independently and observables in present universe such as the CMB
angular power spectrum and the matter power spectrum are predicted by
initial amplitude of each mode and their correlation.

In this paper we adopt the definition of initial perturbations for the
structure formation in \cite{Bucher:1999re}.  We use $X_I(\mathbf{k})$
for representing the initial perturbation of each mode.
\begin{eqnarray}
 \label{eq:initial_modes}
  X_I(\mathbf{k})=\left\{ 
        \begin{array}{ll}
	 \zeta & (\mbox{for AD})\\
	 \mathcal{S}_\mathrm{CDM} & (\mbox{for CI})\\
	 \mathcal{S}_b & (\mbox{for BI})\\
	 \frac{3}{4(1-f_\nu)}\mathcal{S}_\nu & (\mbox{for NID})\\
	 \frac{1}{1-f_\nu}\mathcal{V}_\nu & (\mbox{for NIV})
	\end{array}\right.
\end{eqnarray}
The right hand side of Eq.~(\ref{eq:initial_modes}) is evaluated at the
beginning of the structure formation.  $\zeta$ is the gauge invariant
curvature perturbation and $\mathcal{S}_\mathrm{CDM}$,
$\mathcal{S}_\mathrm{b}$ and $\mathcal{S}_\mathrm{\nu}$ are the entropy
perturbations of CDM, baryon and neutrino, separately.
$\mathcal{V}_\nu=V_\nu-V_\gamma$ is the relative velocity perturbation
of neutrino to photon.  $f_\nu$ is the fraction of the neutrino species
in the energy density of the radiations.  For more detailed definition
of each isocurvature mode, we refer to \cite{Bucher:1999re}.

When we investigate observational constraints on various isocurvature
modes in section~\ref{sec:constraints}, we consider only CI, NID and NIV
modes.  This is because the contribution of CDM and baryon isocurvature
perturbations are brought together into isocurvature perturbations of
matter
\begin{equation}
 \mathcal{S}_m=\frac{\Omega_\mathrm{CDM}}{\Omega_m}
  \mathcal{S}_\mathrm{CDM}
  +\frac{\Omega_b}{\Omega_m}\mathcal{S}_b.
\end{equation}
Thus, the constraint on $\mathcal{S}_b$ is easily obtained from that on
$\mathcal{S}_\mathrm{CDM}$.  

The auto and cross power spectra $\mathcal{P}_{ab}(k)$ of primordial
perturbations can be written as
\begin{equation}
 \mathcal{P}_{IJ}(k)\delta_{\mathbf{k}\mathbf{k}^\prime}=
  \frac{k^3}{2\pi^2}
  \langle X_I(\mathbf{k})^* X_J(\mathbf{k}^\prime)\rangle .
\end{equation}
Here subscripts $I$ and $J$ represent the adiabatic (AD) and four
isocurvature modes (CI, BI, NID and NIV).  We assume power spectra can
be approximated as power-law:
\begin{equation}
   \mathcal{P}_{IJ}(k)
    =A_{IJ}\left(\frac{k}{k_0}\right)^{n_{IJ}-1},
    \label{eq:powerspectra}
\end{equation}
where $k_0$ is a pivot scale and we consistently take $k=0.05$ Mpc$^{-1}$
in the rest of this paper.

Throughout this paper we consider models with only one isocurvature mode
besides adiabatic and tensor modes. This simplification enables us to
capture what models are plausible to generate the initial fluctuations
in the universe including isocurvature and tensor modes and the
resultant constraints on isocurvature and tensor perturbations can be
applied to many theoretical models based on inflation scenarios and
particle physics.  Since we have known that the primordial perturbations
mainly consist of adiabatic perturbations, it is convenient to
normalize the amplitudes of power spectra by the amplitude of the auto
power spectrum of the adiabatic mode, $A_\mathrm{AD}$. Thus, we
parametrize initial power spectra for the scalar perturbations as
\begin{equation}
   A_{IJ}=A_\mathrm{AD}\left(
    \begin{array}{cc}
     1& B_a\cos\theta_a \\
     B_a\cos\theta_a & B_a^2
    \end{array}\right) ,
\end{equation}
where 
\begin{eqnarray}
    B_a&\equiv& \sqrt{A_{aa}/A_\mathrm{AD}} 
     \label{eq:B_a} , \\
    \cos\theta_a&=&A_{\mathrm{AD},a}/\sqrt{A_{aa}A_\mathrm{AD}}.
     \label{eq:cos_theta_a} 
\end{eqnarray}
Subscripts $a$ represent some isocurvature mode being considered (CI, 
NID or NIV).

The tensor perturbations are also generated in inflation models.  When
we refer to tensor to scalar ratio $r$, we usually consider cases that
the scalar perturbation is purely adiabatic.  Since we are considering
isocurvature perturbations along with the adiabatic perturbation here,
we redefine $r$ as 'tensor to adiabatic ratio'.  The power spectrum of
tensor perturbations $\mathcal{P}_g(k)$ is written as follows:
\begin{eqnarray}
   \mathcal{P}_g(k)&=&A_g\left(\frac{k}{k_0}\right)^{n_g} , \\
     &=&rA_\mathrm{AD}\left(\frac{k}{k_0}\right)^{n_g} ,
\end{eqnarray}
where $n_g$ is the spectral index of the tensor perturbation.

Since each mode evolves independently within the framework of the linear
perturbation theory, we can decompose perturbations of the fluids by
initial modes.  As for CMB, the brightness function of photon $\Theta_l$
is written as
\begin{equation}
   \Theta_l(\mathbf{k},\eta)=\sum_I \Theta^I_l(\mathbf{k},\eta),
\end{equation}
where $\Theta^I_l$ are brightness functions which evolve from different
initial perturbation modes.  We introduce a transfer function of photon
for each mode $T^I_l(k, \eta)$:
\begin{equation}
   \Theta^I_l(\mathbf{k}, \eta)=T^I_l(k, \eta)X_I(\mathbf{k}) .
\end{equation}
Then we obtain the angular power spectra of CMB $C_l$,
\begin{eqnarray}
   C_l&=&\sum_{I,J} C_l^{IJ} , \\ 
      &=&A_{AD}
       \left[\hat{C}^\mathrm{adi}_l
	+2B_a\cos\theta_a\hat{C}^\mathrm{cor}_l
	+B_a^2\hat{C}^\mathrm{iso}_l
	+r\hat{C}^\mathrm{tens}_l
	 \right] \label{eq:general_Cl}
\end{eqnarray}
where $\hat{C}_l$'s are the angular power spectra in cases that the
amplitudes of the initial perturbations $A_{IJ}$ are set to be unity and
subscripts $adi$, $iso$, $cor$ and $tens$ represent adiabatic auto,
isocurvature auto, adiabatic-isocurvature cross and tensor power
spectra, respectively. $\hat{C}_l$'s are given by
\begin{eqnarray}
   \hat{C}_l^\mathrm{adi}&=&
    \frac{4\pi}{2l+1}\int\frac{dk}{k}
    \left(\frac{k}{k_0}\right)^{n_\mathrm{AD,AD}-1}
    T^\mathrm{AD}_l(k)^2 ,\\
   \hat{C}_l^\mathrm{iso}&=&
    \frac{4\pi}{2l+1}\int\frac{dk}{k}
    \left(\frac{k}{k_0}\right)^{n_{a,a}-1}
    T^a_l(k)^2 , \\
   \hat{C}_l^\mathrm{cor}&=&
    \frac{4\pi}{2l+1}\int\frac{dk}{k}
    \left(\frac{k}{k_0}\right)^{n_{\mathrm{AD},a}-1}
    T^\mathrm{AD}_l(k)T^a_l(k).
\end{eqnarray}
As for the matter power spectrum,  $P(k)$ can be written in
the same way, 
\begin{equation}
   P(k)=A_{AD}
    \left[
     \hat{P}^\mathrm{adi}(k)
     +2B_a\cos\theta_a\hat{P}^\mathrm{cor}(k)+
     B_a^2\hat{P}^\mathrm{iso}(k)
	      \right], \label{eq:general_Pk}
\end{equation}
where the hatted power spectra $\hat{P}(k)$'s are auto and cross power
spectra with $A_{IJ}$ being unity.

\section{Isocurvature perturbation  based on inflation} \label{sec:models}

So far we have considered generic models with isocurvature and tensor
perturbations.  In this section we consider a model with isocurvature
and tensor perturbations based on inflation models.  We assume there are
two scalar perturbations generated during inflation. One is a curvature
perturbation $\zeta_*$ and the other is a isocurvature perturbation
$\mathcal{S}_*$.  The curvature perturbation $\zeta_*$ raises only
adiabatic mode at the beginning of structure formation, whereas the
isocurvature perturbation $\mathcal{S}_*$ can generally produce both
adiabatic and isocurvature perturbations at the beginning of the
structure formation.  Therefore we can write
\begin{equation}
  \left(\begin{array}{c}
         \zeta\\
	 \mathcal{S}_a
        \end{array}\right)
  =\left(\begin{array}{cc}
    \mathcal{T}_{\zeta, \zeta_*} 
     & \mathcal{T}_{\zeta, \mathcal{S}_*}\\
	  0 
     & \mathcal{T}_{\mathcal{S}_a, \mathcal{S}_*}
   \end{array}\right)
   \left(\begin{array}{c}
       \zeta_*\\
       \mathcal{S}_*
    \end{array}\right)
\end{equation}
Here in the left hand side $\zeta$ and $\mathcal{S}_a$ are initial
curvature and isocurvature perturbations for structure formation,
separately.  $\mathcal{T}$'s are transfer functions which represent how
initial perturbations for structure formation are generated from
perturbations during inflation.  Since the curvature perturbation at
over-horizon scale stays constant in the absence of isocurvature
perturbations, $\mathcal{T}_{\zeta, \zeta_*}=1$.  On the other hand
$\mathcal{T}_{\zeta,\mathcal{S}_*}$ and
$\mathcal{T}_{\mathcal{S}_a,\mathcal{S}_*}$ depend on models.  Then the
initial power spectra for $\zeta$ and $\mathcal{S}_a$ are given by
\begin{eqnarray}
 \mathcal{P}_\mathrm{AD, AD}(k)&=&
  \mathcal{T}_{\zeta, \zeta_*}(k)^2\mathcal{P}_{\zeta_*}(k)+
  \mathcal{T}_{\zeta, \mathcal{S}_*}(k)^2\mathcal{P}_{\mathcal{S}_*}(k) 
  \label{eq:sp_adi} ,\\[0.6em]
 \mathcal{P}_{\mathrm{AD}, a}(k)&=&
  \mathcal{T}_{\zeta, \mathcal{S}_*}(k)
  \mathcal{T}_{\mathcal{S}_a, \mathcal{S}_*}(k)
  \mathcal{P}_{\mathcal{S}_*}(k) \label{eq:sp_cor} ,\\[0.6em]
 \mathcal{P}_{a, a}(k)&=&\mathcal{T}_{\mathcal{S}_a, \mathcal{S}_*}(k)^2
  \mathcal{P}_{\mathcal{S}_*}(k) \label{eq:sp_iso}
\end{eqnarray}
Generally $\zeta_*$ and $\mathcal{S}_*$ may be correlated, such in the
case of multi-field inflation
models~\cite{Bartolo:2001rt,Byrnes:2006fr}.  However, we assume that
perturbations $\zeta_*$ and $\mathcal{S}_*$ are uncorrelated in this
paper. Furthermore, we take power-law forms for the terms in the right
hand sides of Eqs.~(\ref{eq:sp_adi})-(\ref{eq:sp_iso}) as
\begin{eqnarray}
 \mathcal{T}_{\zeta, \zeta_*}(k)^2\mathcal{P}_{\zeta_*}(k)&=&
  A_\mathrm{adi1}\left(\frac{k}{k_0}\right)^{n_\mathrm{adi1}-1} ,\\
 \mathcal{T}_{\zeta, \mathcal{S}_*}(k)^2
  \mathcal{P}_{\mathcal{S}_*}(k)&=&
  A_\mathrm{adi2}\left(\frac{k}{k_0}\right)^{n_\mathrm{adi2}-1} ,\\
 \mathcal{T}_{\zeta, \mathcal{S}_*}(k)
  \mathcal{T}_{\mathcal{S}_a, \mathcal{S}_*}(k)
  \mathcal{P}_{\mathcal{S}_*}(k)&=&
  A_\mathrm{cor}\left(\frac{k}{k_0}\right)^{n_\mathrm{cor}-1} ,
  \label{eq:cor_sp}\\ 
 \mathcal{T}_{\mathcal{S}_a, \mathcal{S}_*}(k)^2
  \mathcal{P}_{\mathcal{S}_*}(k)&=&
  A_\mathrm{iso}\left(\frac{k}{k_0}\right)^{n_\mathrm{iso}-1},
\end{eqnarray}
where $A_\mathrm{cor}$ and $n_\mathrm{cor}$ are given by
\begin{eqnarray}
 A_\mathrm{cor}&=&\pm\sqrt{A_\mathrm{adi2}A_\mathrm{iso}} ,\\
 n_\mathrm{cor}&=&\frac{n_\mathrm{adi2}+n_\mathrm{iso}}{2}.
\end{eqnarray}
Here the sign in the right hand side of the first line comes from a
factor $\mathcal{T}_{\zeta,\mathcal{S}_*}
\mathcal{T}_{\mathcal{S}_a,\mathcal{S}_*}$ in Eq.~(\ref{eq:cor_sp}),
which can be either positive or negative.
Furthermore, we use following parametrizations:
\begin{eqnarray}
 A_\mathrm{AD}&=&A_\mathrm{adi1}+A_\mathrm{adi2} , \\
 B_a^2&=&A_\mathrm{iso}/A_\mathrm{AD} , \\
 B_a\cos\theta_a&=&A_\mathrm{cor}/A_\mathrm{AD} .
\end{eqnarray}

Assuming single-field slow-roll inflation and $\mathcal{S}_*$ is the
isocurvature perturbation for some scalar field ($\neq$ inflaton) whose
mass is negligibly light compared with the Hubble parameter during
inflation, the following inflation consistency relations should be
satisfied:
\begin{equation}
 n_\mathrm{adi2}-1=n_g=-\frac{A_g}{8A_\mathrm{adi1}}
  =-\frac{r}{8\sin^2\theta_a}. 
  \label{eq:consistency}
\end{equation}

We finally obtain the power spectra for CMB and matter, 
\begin{eqnarray}
 C_l&=&A_\mathrm{AD}\left[\sin^2\theta_a\hat{C}_l^\mathrm{adi1}+
		     \cos^2\theta_a\hat{C}_l^\mathrm{adi2}+
		     B_a\cos\theta_a\hat{C}_l^\mathrm{cor}+
		     B_a^2\hat{C}_l^\mathrm{iso}
		     +r\hat{C}_l^\mathrm{tens}\right]\\
 P(k)&=&A_\mathrm{AD}\left[
		      \sin^2\theta_a\hat{P}_l^\mathrm{adi1}+
		      \cos^2\theta_a\hat{P}_l^\mathrm{adi2}+
		      B_a\cos\theta_a\hat{P}^\mathrm{cor}(k)+
		      B_a^2\hat{P}^\mathrm{iso}(k)
	\right],
\end{eqnarray}
Note that there are two terms for adiabatic modes in each $C_l$ and $P(k)$.
CMB and matter power spectra with subscript $adi1$ come from
$\zeta_*$ and those with subscript $adi2$ come from $\mathcal{S}_*$.

\section{Analysis method}
\label{sec:method}
We consider the flat $\Lambda$CDM model, and take the standard value
3.04 for massless neutrino species. We do not consider runnings in the
spectral indices for scalar and tensor perturbations.

Since we are considering isocurvature and tensor perturbations, there
exist six extra parameters $(B_a,\ \cos\theta_a,\ n_{\mathrm{AD},a},\
n_{a,a},\ r,\ n_g)$  or 
$(B_a,\ \cos\theta_a,\ n_\mathrm{adi2},\ n_\mathrm{iso},\ r,\ n_g)$ 
that are absent for a purely adiabatic case.  However, for obtaining
sensible constraints from the present cosmological observations, it is
not suitable to treat all these parameters as free parameters.  In this
paper we adopt some simplifications and fix the spectral indices
$(n_{\mathrm{AD},a},\ n_{a,a}, n_g)$  to some values.
For the case considered in the previous section, we adopt the
inflation consistency relation Eq.~(\ref{eq:consistency}) to fix
$(n_\mathrm{adi2},\ n_\mathrm{iso},\ n_g)$.  These simplifications
reduce extra parameters to three primary free parameters $(B_a,\
\cos\theta_a,\ r)$.

Thus our models have the following nine primary parameters:
\begin{equation}
   (\omega_b, \omega_\mathrm{CDM}, \theta_\mathrm{sound},
    \tau, A_\mathrm{AD},n_{adi}, B_a, \cos\theta_a, r).
\end{equation}
We investigate three models separately depending on the correlation of
isocurvature modes; 1) uncorrelated, 2) totally correlated and 3)
generally correlated isocurvature models. 
When we investigate
uncorrelated $(\cos\theta_a=0)$ and totally correlated isocurvature
$(\cos\theta_a=\pm1)$ models, we fix $\cos\theta_a$ and when we
investigate generally correlated isocurvature modes, we assign flat
prior probabilities on the $\cos\theta_a$ in the range $[-1,1]$.

The likelihood of a model is assessed using the WMAP three-year (WMAP3)
data and likelihood code \cite{Hinshaw:2006ia,Page:2006hz} and SDSS data
release 4 luminous red galaxy sample (SDSS DR4 LRG)
\cite{Tegmark:2006az}.  We include the nonlinear corrections for the
matter power spectrum \cite{Cole:2005sx}, and analytically marginalize
over a bias parameter $b$ and a parameter for nonlinear correction
$Q_\mathrm{nl}$.  We modify the CAMB code \cite{Lewis:1999bs} to
generate CMB and matter power spectra.  Likelihood surfaces are explored
by Markov Chain Monte Carlo methods using CosmoMC \cite{Lewis:2002ah}.
We generate six chains for each model with isocurvature modes and their
correlation and apply the Gelman and Rubin convergence test
\cite{Gelman:1992}.  We finally obtain at least 150,000 samples for each
model, and in some cases over 400,000 samples.

\section{Constraints on isocurvature and tensor perturbations}
\label{sec:constraints}

\subsection{Constraints on the uncorrelated isocurvature models}
\label{subsec:uncorrelated}

Firstly we investigate uncorrelated isocurvature models
($\cos\theta_a=0$).  For the uncorrelated isocurvature models with
tensor mode we impose inflation consistency relation
\begin{equation}
   n_g=-r/8,
\end{equation}
which is realized  in a single-field slow-roll inflation model\footnote{
By single-field inflation model we mean inflation model where the vacuum
energy is determined by a single field and does not depend on other
light fields.
}.

\begin{figure}[t]
 \begin{center}
  \begin{tabular}{ccc}
   CI  &  NID  &  NIV\\
   \includegraphics[scale=1.1]{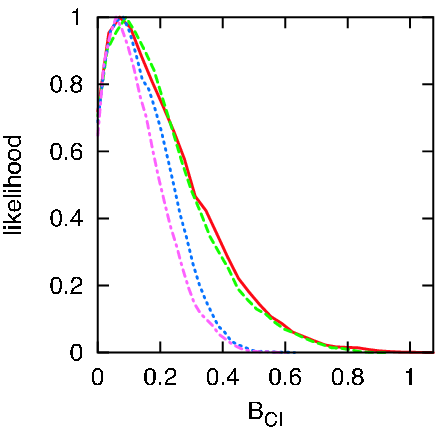}
     &
       \includegraphics[scale=1.1]{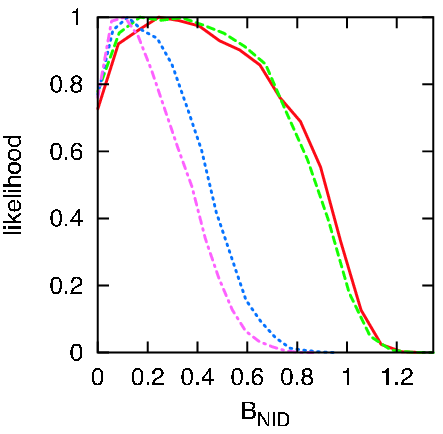} 
     &
	   \includegraphics[scale=1.1]{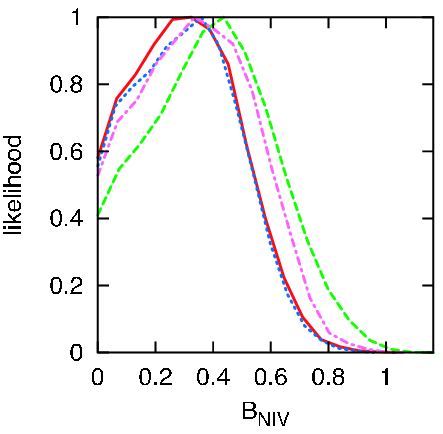}
  \end{tabular}
  \caption{1-dimensional likelihood distributions for uncorrelated
  isocurvature models with CI (left), NID (middle) and NIV (right) mode,
  respectively.  In each panel we show the distributions for models without
  tensor modes using WMAP3 data only (red full), with tensor modes using
  WMAP3 data only (green dashed), without tensor modes using WMAP3 data
  combined with SDSS DR4 LRG data (blue dotted), with tensor modes using
  WMAP3 data combined with SDSS DR4 LRG data (magenta dot-dashed).  }
  \label{fig:UC_b}
 \end{center}
\end{figure} 

\begin{figure}[h]
 \begin{center}
 \begin{tabular}{ccc}
  CI & NID & NIV \\
  \includegraphics[scale=1.1]{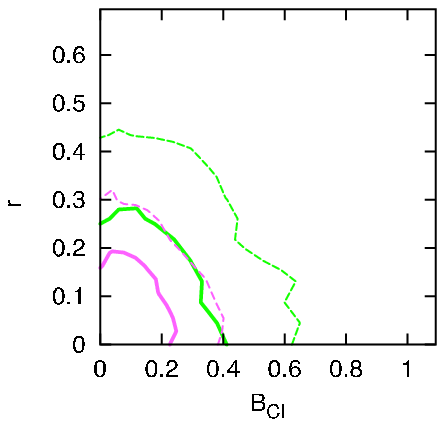}  &
      \includegraphics[scale=1.1]{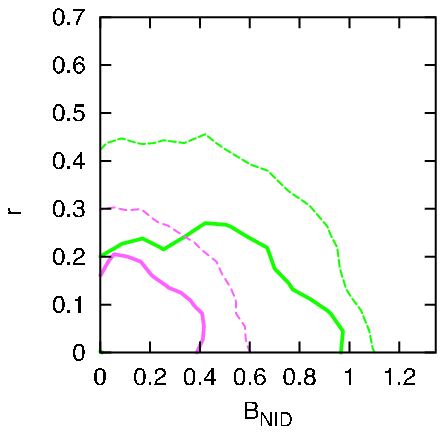} &
	 \includegraphics[scale=1.1]{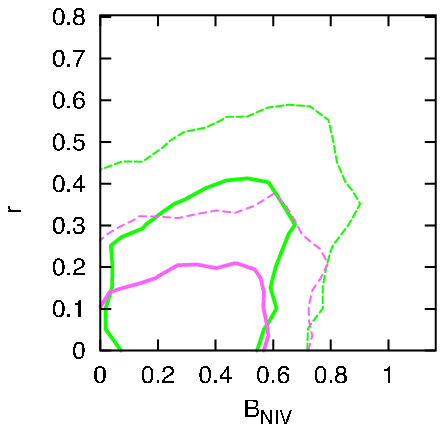}
 \end{tabular}
  \caption{68 \% (full) and 95\% (dashed) 2-dimensional constraints on
  the uncorrelated isocurvature models CI (left), NID (middle) and NIV
  (left) with tensor modes. We present constraints using WMAP3 data only
  (green) and combined with SDSS DR4 LRG data.  } \label{fig:UC_b-r}
 \end{center}
\end{figure} 

We present 1d-marginalized likelihood distributions for the uncorrelated
isocurvature models in Figure~\ref{fig:UC_b}.  We also show 95\%
confidence limits (c.l.) on $B_a$ and $r$ for each models with (without)
tensor modes from the combination of WMAP3 and SDSS DR4 LRG data in
Table~\ref{tbl:UC}.  The CDM isocurvature (CI) and neutrino isocurvature
density (NID) modes are rather tightly constrained and there is no
improvement in minimum $\chi^2$. On the other hand, presence of neutrino
isocurvature velocity (NIV) modes tends to be favored by the present CMB
and LSS data, though not yet at decisive level.  Thus, we find no
statistical support for finite contribution from uncorrelated
isocurvature modes and CMB and LSS data are consistent with purely
adiabatic initial scalar perturbations.

We also find 95\% limits on tensor modes.  Comparing with the constraint
$r\le0.30$ (95\% c.l.) for the model with purely adiabatic scalar
perturbations~\cite{Tegmark:2006az}, we find that the upper limits on
$r$ for models with uncorrelated isocurvature are roughly same as that
for the purely adiabatic model.  This is because uncorrelated the
isocurvature modes (except for uncorrelated NIV mode) and tensor mode
contribute to the large scale anisotropy of CMB positively and there are
no parameter degeneracy.  For uncorrelated NIV mode, situations are
little different since CMB power spectrum for NIV mode is relatively
similar to that for AD mode [ see, e.g.,
Fig.~1. in~\cite{Bucher:2000cd}].  Thus, the upper limit on $r$ for
uncorrelated NIV models is higher than those for other isocurvature
models but it is still comparable with that for the purely adiabatic
model.

\begin{table}[h]
 \begin{center}
  \begin{tabular}{|r||r|r|r|}\hline
   & CI & NID &NIV \\ \hline
   $B_a\le$ & 0.31(0.33) & 0.51(0.54) & 0.69(0.62) \\ 
   $r\le$ & 0.26 & 0.25 & 0.31 \\ \hline
   $\Delta\chi^2_\mathrm{min}$&0(0)&0(0)&-1(-1)\\ \hline
  \end{tabular}
  \caption{Constraints on $B_a$ and $r$ at 95\% c.l. for uncorrelated
  isocurvature models with tensor modes (without tensor modes) from
  WMAP3+SDSS DR4 LRG.  We also show the changes of the minimum $\chi^2$
  values from the purely adiabatic model.}  \label{tbl:UC}
 \end{center}
 \end{table}

It is known there are some parameter degeneracies among fractions of
isocurvature modes, $B_a$, and other cosmological parameters $\omega_b$,
$\omega_\mathrm{CDM}$ and $n_\mathrm{adi}$.  These degeneracies are
understood by recognizing that the constraints on isocurvature modes
rely mainly on the angular scale and the peak hight of the first
acoustic peak in the CMB TT power spectrum. The relative hight of the
first acoustic peak to the anisotropy at large angular scale, increases
as the baryon density increases through compressions of the
photon-baryon fluid. It also increases as the CDM density decreases and
the early integrated Sachs-Wolfe effect is enhanced.  Finally the peak
height increases as the spectral index increases which leads to larger
primordial fluctuations in small scales.  For CI and NID modes, increase
in $B_a$ decreases the relative peak hight of the acoustic peak. Thus
some cancelations exist among $B_a$, $\omega_b$, $\omega_\mathrm{CDM}$
and $n_s$ and parameter degeneracies arise.  But for NIV mode, increase
in $B_a$ does not decrease the peak hight much and parameter
degeneracies in NIV models are weaker than in uncorrelated CI and NID
models.  Though the peak hight has also strong dependence of the optical
depth $\tau$, the polarization power spactra (TE and EE) of WMAP3
constrains $\tau$ tightly and no parameter degeneracy between $B_a$ and
$\tau$ is seen.  Some parameter degeneracies such as degeneracy between
$B_a$ and $\omega_\mathrm{CDM}$ are broken by inclusion of LSS data 
and the constraints on $B_a$ improve  (for CI and NID modes seen in
Figure~\ref{fig:UC_b}).

\subsection{Constraints on totally correlated isocurvature models}
\label{subsec:totally_correlated}

Next we investigate totally correlated isocurvature models
($\cos\theta_a=\pm1$). Firstly we define parameters for collecting both
positively ($\cos\theta_a=1$) and negatively ($\cos\theta_a=-1$)
correlated isocurvature models as 
\begin{eqnarray}
   B_a^\prime&=&B_a\cos\theta_a
   = \left\{\begin{array}{cc}
       B_a & (\mbox{for } \cos\theta_a=1) \label{eq:Bprime}\\
       -B_a & (\mbox{for }\cos\theta_a=-1)
      \end{array}\right. .
\end{eqnarray}
$B_a^\prime$ take either positive, 0 and negative values.  For totally
correlated isocurvature models with tensor modes we assume scale
invariant tensor perturbations $n_g=1$.

\begin{figure}[t]
 \begin{center}
  \begin{tabular}{ccc}
   CI&NID&NIV
   \\
   \includegraphics[scale=1.1]{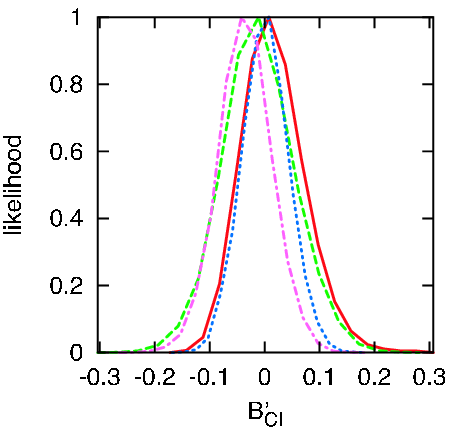} &
   \includegraphics[scale=1.1]{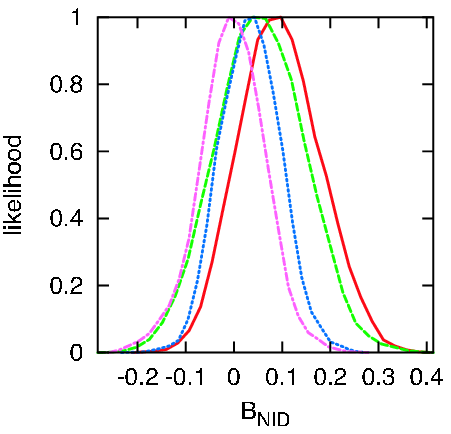}&
	   \includegraphics[scale=1.1]{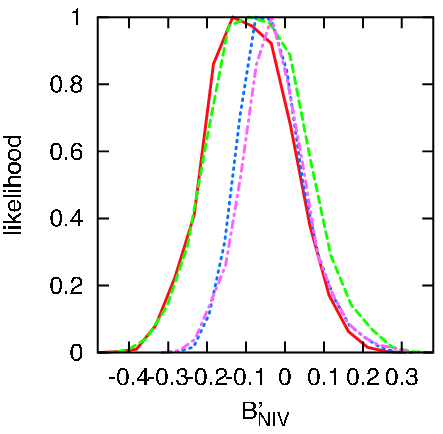}
  \end{tabular}
  \caption{1-dimensional likelihood distributions for totally correlated
  isocurvature models .  Considered isocurvature and tensor modes and
  combinations of data are same as Figure~\ref{fig:TC_b}.}
  \label{fig:TC_b}
 \end{center}
\end{figure} 


\begin{figure}[t]
 \begin{center}
 \begin{tabular}{ccc}
  CI&NID&NIV
  \\
  \includegraphics[scale=1.1]{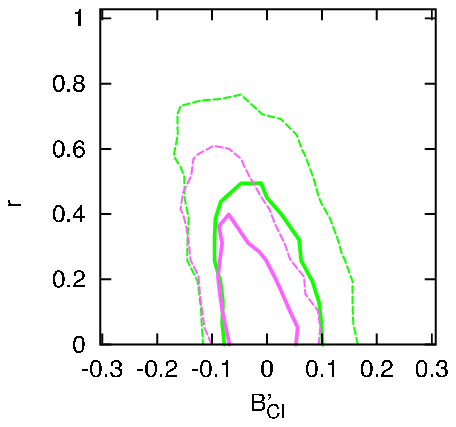}&
      \includegraphics[scale=1.1]{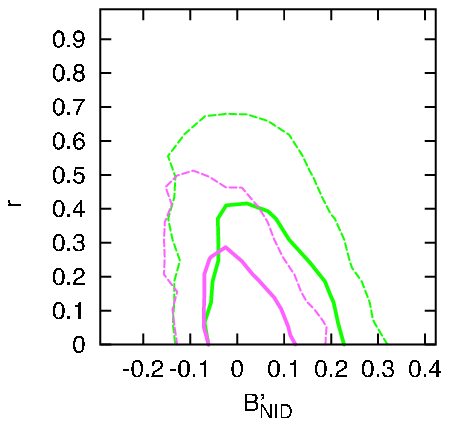} &
	  \includegraphics[scale=1.1]{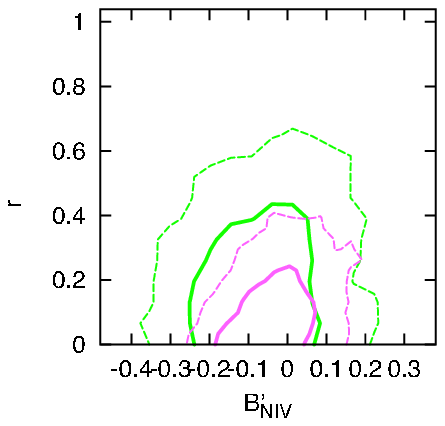}
 \end{tabular}
  \caption{68 \% and 95\% 2-dimensional constraints on the totally
  correlated isocurvature models with tensor modes.  Contours are same
  as Figure~\ref{fig:UC_b-r}.}
  \label{fig:TC_b-r}
 \end{center}
\end{figure} 

We show 1d-marginalized likelihood distributions for totally correlated
isocurvature models in Figure~\ref{fig:TC_b}. 95 \% limits and relative
changes in minimum $\chi^2$ from the purely adiabatic modes are
presented in Table~\ref{tbl:TC}.  We find no improvement in $\chi^2$
values and the present observations of CMB and LSS are consistent with
the purely adiabatic initial conditions.  For any isocurvature modes,
the limits on $B_a$ for totally correlated models are found to be more
stringent than for uncorrelated models.  This is because, for totally
correlated models, the correlation terms $\hat{C}_l^\mathrm{cor}$ and
$\hat{P}^\mathrm{cor}(k)$ give significant contributions to the CMB and
matter power spectra (see Eqs.~(\ref{eq:general_Cl}) and
(\ref{eq:general_Pk})), which is not present for uncorrelated
models. Therefore both CMB and matter power spectra are affected much if
totally correlated isocurvature perturbations are present and limits
becomes more stringent.

On the other hand the constraints on tensor modes $r$ for CI and NID
modes becomes weaker, compared to those for uncorrelated models.  This
is because anti-correlated CI and NID modes decrease the anisotropies in
large angular scales of the CMB TT power spectrum, which can be partly
canceled by contributions from tensor modes.  However, the constraints
on tensor modes are not affected much for correlated NIV mode.


\begin{table}[t]
 \begin{center}
  \begin{tabular}{|r||r|r|r|}\hline
   & CI & NID &NIV \\ \hline
   $B^\prime_a$ 
   & $\le$ 0.056(0.087) & $\le$ 0.118(0.173) & $\le$ 0.130(0.101) \\ 
   & $\ge$ -0.129(-0.080) &  $\ge$ -0.151(-0.090) & $\ge$ -0.189(-0.174) \\ 
   $r\le$ 
   &   0.49 & 0.44 & 0.30 \\ \hline
   $\Delta\chi^2_\mathrm{min}$
   & 0(0) & 0(0) & 0(0)\\ \hline
  \end{tabular}
  \caption{Constraints for totally correlated models with tensor modes
  (without tensor modes) from WMAP3+SDSS DR4 LRG.  We show $B^\prime_a$
  and $r$ at 95 \% c.l. and the changes of the minimum $\chi^2$ values
  from the purely adiabatic model.}  \label{tbl:TC}
 \end{center}
\end{table}

\subsection{Constraints on generally correlated isocurvature models} 
\label{subsec:generally_correlated}
\begin{figure}[htb]
 \begin{center}
  \begin{tabular}{ccc}
   CI&NID&NIV\\
   \includegraphics[scale=1.1]{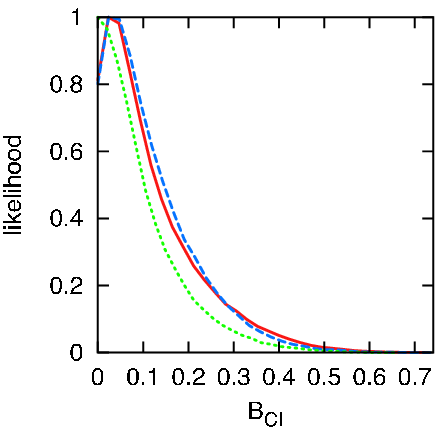} &
       \includegraphics[scale=1.1]{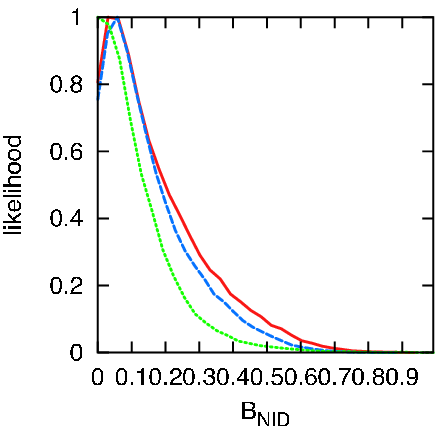} &
	   \includegraphics[scale=1.1]{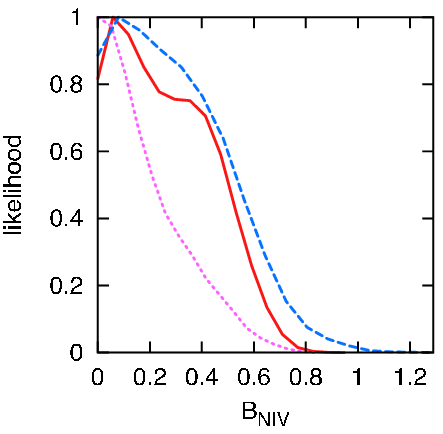}
  \end{tabular}
  \caption{1-dimensional likelihood distributions for generally
  correlated isocurvature models with CI (left), NID (middle) and NIV
  (right) mode.  In each panel, we show distributions for models without
  tensor modes (red full), with tensor modes imposed inflation
  consistency relations (green dotted) and with tensor modes with fixed
  spectral index $n_g=1$ (blue dashed).}  \label{fig:GC_b}
 \end{center}
\end{figure} 

\begin{figure}[h!]
 \begin{center}
  \begin{tabular}{ccc}
   CI&NID&NIV
   \\
   \includegraphics[scale=1.1]{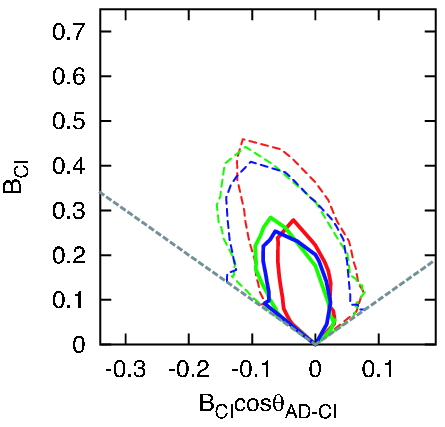} &
       \includegraphics[scale=1.1]{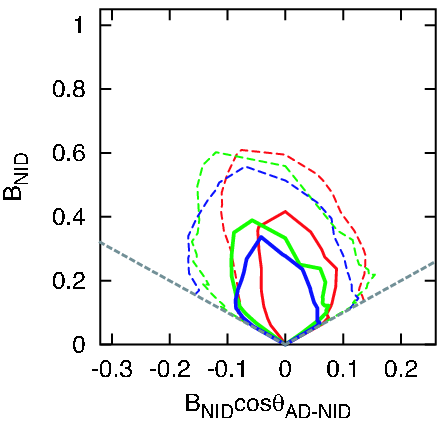} &
	   \includegraphics[scale=1.1]{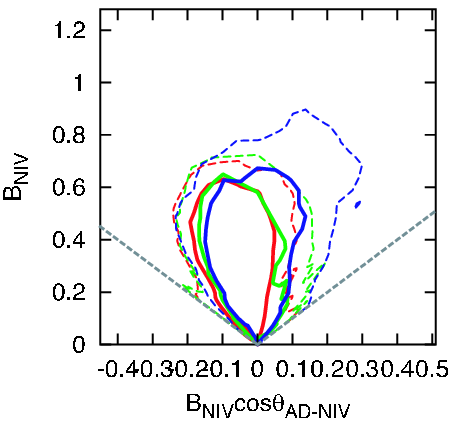}
  \end{tabular}
  \caption{68\% (solid) and 95\% (dashed) 2-dimensional constraints on
  the generally correlated isocurvature models.  We present constraints
  for models with CI (left), NID (middle) and NIV (right) mode,
  separately. In each panel, we show constraints on models without
  tensor modes (red), with tensor modes imposed inflation consistency
  relations (green) and with tensor modes with fixed spectral index
  $n_g=1$ (blue).  Black dashed lines represents $\cos\theta_a=\pm 1$.}
  \label{fig:GC_b-bcos}
 \end{center}
\end{figure} 

\begin{figure}[h!]
 \begin{center}
  \begin{tabular}{ccc}
   CI&NID&NIV
   \\
   \includegraphics[scale=1.1]{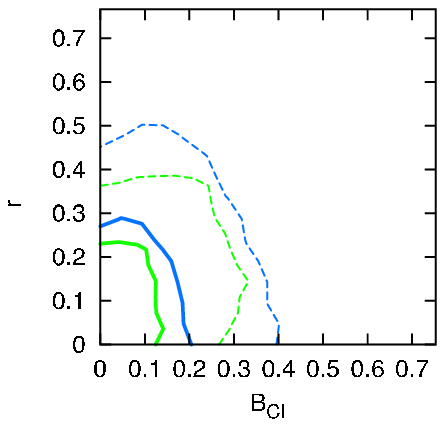} &
       \includegraphics[scale=1.1]{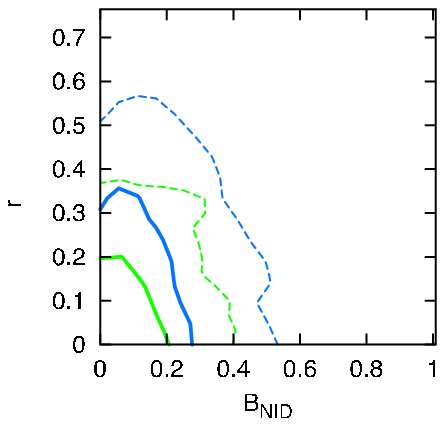} &
	   \includegraphics[scale=1.1]{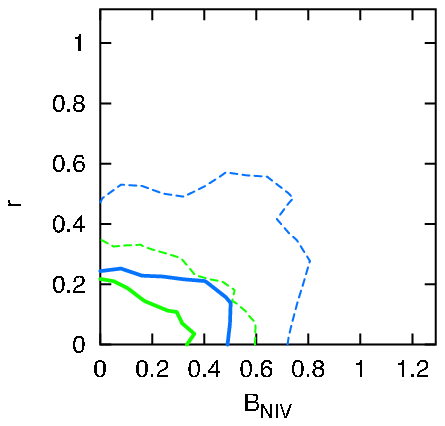}
  \end{tabular}
  \caption{68 \% and 95\% 2-dimensional constraints on the totally
  correlated isocurvature models with tensor modes.  Contours are same
  as Figure~\ref{fig:GC_b-bcos}.}
  \label{fig:GC_b-r}
 \end{center}
\end{figure} 

We finally investigate generally correlated isocurvature models ($-1\le
\cos\theta_a\le1$).  For models with generally correlated isocurvature
modes and tensor modes we consider two versions of models, models
on which the inflation consistency relations Eq.~(\ref{eq:consistency})
is imposed~\footnote{
Since Eq.~(\ref{eq:consistency}) assumes slow-roll inflation, we must
care that samples in MCMC chains should not take large values for
slow-roll parameter $\epsilon=-n_g/2= -r/16(1-\cos^2\theta_a)$. However,
we have checked that $\epsilon$ for each sample takes no more than $0.1$
and our use of Eq.~(\ref{eq:consistency}) is consistent.  This is because
the CMB power spectrum at large angular scale disfavors such negative
large value of $n_g$.}
and models with fixed spectral index $n_g=1$ for tensor modes.

We present the 1d-marginalized likelihood distributions for generally
correlated isocurvature models in Figure~\ref{fig:GC_b}, and 2d
likelihood contours in Figure~\ref{fig:GC_b-bcos}.  We also show 95\%
confidential limits for $B_a$ and $r$, mean values and 68\% confidential
limits for $\cos\theta_a$ and changes in minimum $\chi^2$ values from
the purely adiabatic models in Table~\ref{tbl:GC_TIC}-\ref{tbl:GC_NT}.
Still, we find that observations are consistent with adiabatic initial
conditions.

The upper bounds for $B_a$ are similar to the uncorrelated isocurvature
models.  These results are also guessed from the results obtained in
Section~\ref{subsec:uncorrelated} and \ref{subsec:totally_correlated},
since, as we have seen, the allowed contribution of isocurvature
perturbations are higher for uncorrelated isocurvature models than those
for totally correlated modes.

We also present constraints for isocurvature and tensor perturbations in
Figure~\ref{fig:GC_b-r}.  We can see that when we impose inflation
consistency relations, upper bounds for tensor modes $r$ are roughly
same as those for uncorrelated isocurvature models.  This can be
understood as follows. When correlations of isocurvature perturbations
with adiabatic perturbations are either positively or negatively large
($\cos^2\theta_a\simeq1$), the spectral index for tensor modes $n_g$
takes large negative values for fixed values for $r$, resulting in too
much fluctuations for CMB anisotropy at large angular scales, which is
disfavored from observations. Therefore large $r$ is allowed only when
the correlation of isocurvature perturbations is small and hence the
resulted bounds on $r$ are similar to those for uncorrelated
isocurvature models.  On the other hand, when we take the fixed spectral
index for tensor modes, $n_g=1$, correlations of isocurvature
perturbations can become large and the bounds for $r$ weaken as in the
cases for totally correlated isocurvature and tensor perturbation models
with $n_g=1$.

\begin{table}[ht]
 \begin{center}
  \begin{tabular}{|r||r|r|r|}\hline
   & CI & NID &NIV \\ \hline
   $B_a\le$ 
   & 0.28 & 0.31 & 0.58 \\ 
   $\cos\theta_a$ 
   & $-0.25\pm0.52$ & 
	   $0.03\pm0.62$ & 
	       $0.05\pm0.50$ \\ 
   $r\le$ & 0.32 & 0.28 & 0.31 \\ \hline
   $\Delta\chi^2_\mathrm{min}$&0&0&-2\\ \hline
  \end{tabular}
  \caption{Constraints for generally correlated models with tensor modes
  imposed inflation consistency relations on from WMAP3+SDSS DR4 LRG.
  We show 95\% c.l. for $B_a$ and $r$, mean values and 68\% c.l. for
  $\cos\theta_a$, and changes of the minimum $\chi^2$ values from the
  purely adiabatic model.}  \label{tbl:GC_TIC}
 \end{center}
 \end{table}

\begin{table}[ht]
 \begin{center}
  \begin{tabular}{|r||r|r|r|}\hline
   & CI & NID &NIV \\ \hline
   $B_a\le$ 
   & 0.29 & 0.41 & 0.76 \\ 
   $\cos\theta_a$ 
   & $-0.04\pm0.43$ & 
	   $-0.10\pm0.47$ & 
	       $0.06\pm0.34$ \\ 
   $r\le$ & 0.43 & 0.50 & 0.73 \\ \hline
   $\Delta\chi^2_\mathrm{min}$&0&0&-2\\ \hline
  \end{tabular}
  \caption{Same as Table~\ref{tbl:GC_TIC} except for $n_g=1$.}
  \label{tbl:GC_NIC}
 \end{center}
 \end{table}

\begin{table}[h!]
 \begin{center}
  \begin{tabular}{|r||r|r|r|}\hline
   & CI & NID &NIV \\ \hline
   $B_a\le$ 
   & 0.33 & 0.47 & 0.59 \\ 
   $\cos\theta_a$ 
   & $-0.06\pm0.34$ & 
	   $0.06\pm0.45$ & 
	       $0.14\pm0.40$ \\ \hline
   $\Delta\chi^2_\mathrm{min}$&0&0&-2\\ \hline
  \end{tabular}
  \caption{Constraints for generally correlated models without tensor modes
  from WMAP3+SDSS DR4 LRG. }
  \label{tbl:GC_NT}
 \end{center}
 \end{table}

\section{Application}
\label{sec:application}

In this section we apply the constraints on the isocurvature
perturbation obtained in the previous section to some specific models of
particle cosmology. We investigate two kinds of models, axion
isocurvature perturbation models and curvaton scenarios.

\subsection{Constraints on axion isocurvature perturbation 
and inflation models}
\label{subsec:axion}
Axion, which is originally proposed as a remedy for strong CP problem in
QCD~\cite{Peccei:1977hh,Weinberg:1977ma,Wilczek:1977pj}, is a candidate
for CDM. The properties of axion, such as the decay constant and its
couplings to ordinary matters are constrained from various observations
of astrophysical and cosmological
phenomena~\cite{Turner:1989vc,Raffelt:1990yz}.  In inflationary
universe, the axion field has  CDM isocurvature
perturbations~\cite{
Seckel:1985tj,
Turner:1990uz,
Steinhardt:1983ia,
Kawasaki:1995ta,Kawasaki:1997ct,Kasuya:1997td,Burns:1997ue,Kanazawa:1998pa}
and they are constrained from observations of CMB and
LSS~\cite{Lyth:1989pb,Beltran:2006sq}.  Firstly, we briefly review how
the axion becomes CDM and its isocurvature fluctuation arises in the
early universe.

We consider the case where the PQ symmetry is spontaneously broken when
the universe is at the stage of inflation.  During inflation the
expectation value of the axion field is very smooth but fluctuates by
the amount of the Hubble parameter $H_\mathrm{inf}$.
The mean value of the axion $\chi$ and its fluctuation $\delta\chi$ can
be represented as
\begin{equation}
   \chi=f_a\theta_i,
    \label{eq:mean}
\end{equation}
\begin{equation}
   \delta\chi=H_\mathrm{inf}/2\pi.
    \label{eq:fluctuation}
\end{equation}
Here, $f_a$ is the axion decay constant and $\theta_i$ is the initial
phase of the axion field which takes an arbitrary value between $-\pi$ and
$\pi$.

When the cosmic temperature is much higher than the QCD scale ($T \gg
\Lambda_\mathrm{QCD}$), the axion has no potential and its field value
stays constant. As the universe expands and its temperature decreases,
the universe undergoes the QCD phase transition and the axion obtains
mass which depends on the temperature  $T$ as ~\cite{Gross:1980br} 
\begin{equation}
   m_\chi(T)=\lambda m_\chi(T\!=\!0)
    \left(\frac{T}{\Lambda_\mathrm{QCD}}\right)^{p},
\end{equation}
where $\lambda\simeq 0.1$ and $p\simeq -4$. When the axion mass becomes
equal to the Hubble parameter $[ m_\chi(T) \sim H(T))$ ], the axion field
starts to oscillate.
After the axion starts oscillation its energy density scales as $a^{-3}$
and behaves as CDM. The density parameter of the axion is given by
\begin{equation}
   \omega_\chi\equiv\Omega_\chi h^2=4.3\times \gamma~\theta_i^2
    \left(\frac{\Lambda_\mathrm{QCD}}{200\mbox{MeV}}\right)^{-2/3}
    \left(\frac{m_\chi(T\!=\!0)}{1\mu\mbox{eV}}\right)^{-7/6}, 
    \label{eq:density_axion}
\end{equation}
where $\gamma$ is the dilution factor. If there occurs no entropy
release after the axion starts oscillation, $\gamma=1$.  The mass of the
axion at zero temperature is determined by its decay constant
$f_a$~\cite{Weinberg:1996kr} as
\begin{equation}
   m_\chi(T\!=\!0) = 1.3\times10^{-3}\mbox{eV}
    \left(\frac{f_a}{10^{10}\mbox{GeV}}\right)^{-1}.
    \label{eq:mass_axion}
\end{equation}
Thus Eq.~(\ref{eq:density_axion}) can be rewritten in terms of $f_a$ as 
\begin{equation}
   \omega_\chi=1.0\times10^{-3}\times\gamma\theta_i^2
    \left(\frac{\Lambda_\mathrm{QCD}}{200\mbox{MeV}}\right)^{-2/3}
    \left(\frac{f_a}{10^{10}\mbox{GeV}}\right)^{7/6}
    \label{eq:density_axion2}.
\end{equation}

The axion isocurvature (entropy) perturbation is written as
\begin{equation}
   \mathcal{S}_\chi\equiv \frac{\delta n_\chi}{n_\chi}
    -\frac{\delta n_\gamma}{n_\gamma},
\end{equation}
where $n_\chi$ and $n_\gamma$ are the number densities of axion and
photon, respectively.  Axion isocurvature perturbation is given by the
fluctuation of the axion field during inflation,
\begin{equation}
    \mathcal{S}_\chi=2\frac{\delta\chi}{\chi}
     =\frac{H_\mathrm{inf}}{\pi f_a\theta_i},
\end{equation}
where we have used Eqs.~(\ref{eq:mean}) and (\ref{eq:fluctuation}) at
the second equality.

We consider the general case where CDM consists of axion and other
particles and assume that only axion contributes to isocurvature
perturbation.  Then the CDM isocurvature perturbation are given by
\begin{equation}
   \mathcal{S}_\mathrm{CDM}=
    \frac{\omega_\chi}{\omega_\mathrm{CDM}}\mathcal{S}_\chi.
    \label{eq:S_cdm_ax}
\end{equation}

The curvature perturbation $\zeta$ and tensor perturbations
$h_{+,\times}$ are also generated during inflation and is written as
\begin{eqnarray}
   \zeta&=&-\frac{H_\mathrm{inf}}{d\phi/dt}\delta\phi ,\\
    h_{+,\times}&=&\frac{H_\mathrm{inf}}{\sqrt{2}M_\mathrm{Pl}},
\end{eqnarray}
where $\phi$ is the field value of the inflaton and $M_\mathrm{Pl}\equiv
\sqrt{8\pi G}$ is the reduced Planck mass.
We thus obtain power spectra of the adiabatic, CDM isocurvature and
tensor modes as
\begin{eqnarray}
   A_\mathrm{AD}&=&
    \frac{H_\mathrm{inf}^2}{8\pi^2M_\mathrm{Pl}^2\epsilon} ,
    \label{eq:A_zeta}\\
   A_\mathrm{CI}&=&\frac{\omega_\chi^2}{\omega_\mathrm{CDM}^2}
    \frac{H_\mathrm{inf}^2}{\pi^2f_a^2\theta_i^2} , \label{eq:A_ax}\\
   A_g&=&\frac{H_\mathrm{inf}^2}{6\pi^2M_\mathrm{Pl}^2} , \\
   n_\mathrm{AD}-1&=&-6\epsilon+2\eta , \\
   n_\mathrm{CI}-1&=&n_g=-2\epsilon .
\end{eqnarray}
Here we assume slow roll inflation and $\epsilon$ and $\eta$ are slow
roll parameters given by
\begin{eqnarray}
   \epsilon&=&\frac{1}{2}M_\mathrm{Pl}^2
    \left(\frac{dV/d\phi}{V}\right)^2 ,\\
   \eta&=&M_\mathrm{Pl}^2\frac{d^2V/d\phi^2}{V}.
\end{eqnarray}

Since the axion isocurvature and the curvature perturbations
are uncorrelated, $\cos\theta_\mathrm{CI}=0$. Using
Eq.~(\ref{eq:B_a}) with Eqs.~(\ref{eq:S_cdm_ax}), (\ref{eq:A_zeta}) and
(\ref{eq:A_ax}), we obtain $B_\mathrm{CI}$ and $r$ as 
\begin{eqnarray}
   B_\mathrm{CI}&=& \frac{\omega_\chi}{\omega_\mathrm{CDM}} 
    \frac{2\sqrt{2\epsilon}M_\mathrm{Pl}}{f_a\theta_i},
    \label{eq:B_CI_ax}\\[0.5em]
   r&=&16\epsilon .
\end{eqnarray}

\begin{figure}[t]
 \begin{center}
  \includegraphics[scale=1.8]{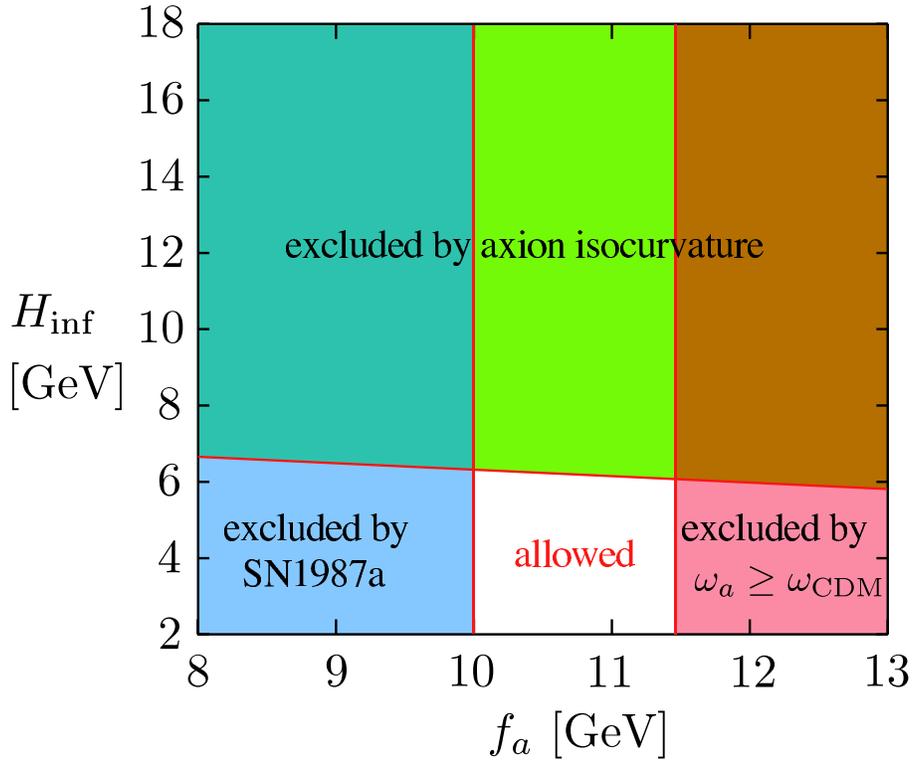}
  \caption{Constraints on the axion decay constant and Hubble parameter
  in the inflation universe. The colored regions are excluded by
  cosmic density of the axion (red),  SN1987A (blue)
  and axion isocurvature perturbation (green).}  \label{fig:fa_Hinf}
 \end{center}
\end{figure} 

We are now ready to study constraints on axion and inflation models.
From now on, we take $\Lambda_\mathrm{QCD}=200$ MeV, $\theta_i=1$ and
assume no entropy release occurs after axion starts oscillation,
i.e. $\gamma=1$.  Firstly we obtain a upper bound on the axion decay
constant $f_a$ from the requirement that the energy density of axion
should not exceed the observed matter density in the present universe,
$\omega_\chi\le\omega_\mathrm{CDM}$.  Combined with the lower bound
obtained from supernovae 1987a~\cite{Yao:2006px} the axion decay
constant should be in following range.
\begin{equation}
   10^{10}\mbox{GeV}\le f_a \le 4.1  \times 10^{11} \mbox{GeV} .
\end{equation}

With using the bound for the CDM isocurvature mode 
in Table~\ref{tbl:UC} we obtain the limits on the inflation parameters as
\begin{eqnarray}
   &H_\mathrm{inf}&\le 10^7\mbox{GeV},\\
   &\epsilon&\le 10^{-16},\\
   -0.05\le&\eta&\le 0.06 .
\end{eqnarray}
We also present obtained bound in the $f_a$-$H_\mathrm{inf}$ plane in
Figure~\ref{fig:fa_Hinf}.  From Eqs.~(\ref{eq:A_zeta}),
(\ref{eq:density_axion}) and (\ref{eq:B_CI_ax}) the ratio of the CDM
isocurvature perturbation to the adiabatic one $B_\mathrm{CI}$ is
written as
\begin{equation}
   B_\mathrm{CI}=6.9\times10^{-2} 
    \left(\frac{\omega_\mathrm{CDM}}{0.1}\right)^{-1}
    \left(\frac{A_\mathrm{AD}}{2.1\times10^{-9}}\right)^{1/2}
    \left(\frac{H_\mathrm{inf}}{10^7\mbox{GeV}}\right)
    \left(\frac{f_a}{10^{10}\mbox{GeV}}\right)^{1/6}
    \propto H_\mathrm{inf}f_a^{1/6}.
\end{equation}
Therefore the upper bound on the Hubble parameter $H_\mathrm{inf}$
during inflation becomes lower as $f_a$ takes larger value.

The resultant constraints on $f_a$ and $H_\mathrm{inf}$ are comparable
with those in \cite{Beltran:2006sq}, where the authors used the
constraints on isocurvature and tensor modes derived by considering
models with either of them, not both.  We have analyzed models with both
isocurvature and tensor modes but constraints on $H_\mathrm{inf}$
have not improved much. This is because the axion model predicts much
less tensor perturbation than isocurvature one since $f_a\ll
M_\mathrm{Pl}$.  Thus, the obtained constraints on $f_a$ and
$H_\mathrm{inf}$ do not change by inclusion of tensor modes.  We can
say oppositely that if nonzero contributions of tensor modes are
suggested by future observations, the axion isocurvature model will be
completely excluded.

In the case where the initial misalignment of axion field is
accidentally much smaller than its natural value, $\theta_i\ll 1$, the
constraints on the axion decay constant weakens since initial amplitude
for the oscillation of axion field becomes smaller.  
\begin{equation}
   \omega_\chi\propto f_a^{7/6}\theta_i^2 .\\
\end{equation}
The constraints on $H_\mathrm{inf}$ also weakens.  Although the
amplitude of the isocurvature perturbation in axion field becomes larger
by decrease of initial misalignment, however, its fraction in CDM
isocurvature perturbation becomes smaller since the fractions of axion
in CDM becomes lower.
\begin{equation}
   B_\mathrm{CI}\propto f_a^{1/6}\theta_i H_\mathrm{inf} .
\end{equation}

As we stated in the early part of this section, we have so far
considered the case where the PQ symmetry is spontaneously broken during
inflation.  When the PQ symmetry is not broken during inflation or
restored by the reheating after inflation, the inflation scale
$H_\mathrm{inf}$ is not bounded by the constraints on the CDM
isocurvature perturbations.

\subsection{Constraints on curvaton models}
\label{subsec:curvaton}

In curvaton scenarios curvature perturbations are generated from the
fluctuation of a scalar field ($=$ curvaton) which is isocurvature at
the epoch of inflation.  We firstly briefly review curvaton scenarios
and then apply the constraints obtained in the previous section to them.

We represent a curvaton field as $\sigma$ and an inflaton field as
$\phi$.  Here we consider the case that the curvaton field is
sufficiently light compared with the Hubble parameter during
inflation. Then the mean value and fluctuation of the curvaton field are
given as
\begin{eqnarray}
    \sigma&=&\sigma_i ,\\
    \delta\sigma&=&\frac{H_\mathrm{inf}}{2\pi}
\end{eqnarray}
We represents the curvature perturbation generated during inflation as
$\zeta_*$ and the isocurvature perturbation of the curvaton field as
$\mathcal{S}_\sigma=2\delta\sigma/\sigma_i$.

Until the Hubble parameter of the universe becomes below the mass of the
curvaton mass, the expectation value of the curvaton field is
constant. After the Hubble parameter becomes comparable to the mass of
the curvaton, the curvaton field starts oscillation and its energy
dominates the universe.
When the curvaton starts dominating the density of the universe, its
fluctuation generates the curvature perturbation. After the curvaton 
decays, its energy turns into the radiation. If the curvaton produce the
CDM, baryon or lepton number, their fluctuations also obey the
fluctuation of the curvaton before its decay.  Then various perturbations
that are relevant for the structure formation are given  by
\begin{eqnarray}
    \zeta&=&\zeta_*+\frac{1}{3}\mathcal{S}_\sigma , \\[0.5em]
    \mathcal{S}_\mathrm{CDM}&=&
     (r_\mathrm{CDM}-1)\mathcal{S}_\sigma+f_\nu\mathcal{S}_\nu 
     \label{eq:cur_cdm} , \\[0.5em]
    \mathcal{S}_b&=&(r_\mathrm{B}-1)\mathcal{S}_\sigma
     +f_\nu\mathcal{S}_\nu 
     \label{eq:cur_b} , \\[0.5em]
    \mathcal{S}_\nu&=&
     \frac{45}{7}\left(\frac{\xi}{\pi}\right)^2
     (r_\mathrm{L}-1)\mathcal{S}_\sigma, 
     \label{eq:cur_nu}
\end{eqnarray}
where $r_\mathrm{CDM}$, $r_\mathrm{B}$, $r_\mathrm{L}$ are the fractions
of the CDM, baryon number and lepton number densities produced by or
after the decay of the curvaton in the present densities.  $\xi$ is the
neutrino asymmetry parameter and we keep only the leading term of order
in $\xi/\pi$ in Eq.~(\ref{eq:cur_nu}) since $\xi$ is constrained from
Big Bang Nucleosynthesis (BBN)\cite{Serpico:2005bc} using the the
observed helium abundance in~\cite{Olive:2004kq} as
\begin{equation}
   |\xi|\le0.07.
\end{equation}

The neutrino isocurvature density perturbation $\mathcal{S}_\nu$ in
Eq.~(\ref{eq:cur_nu}) comes from the isocurvature perturbation in lepton
number density $\mathcal{S}_L\equiv(\delta n_\mathrm{L}/n_\mathrm{L}-
\delta n_\gamma/n_\gamma)$ \cite{Lyth:2002my}.  This is because nonzero
lepton number density in the universe $n_L\ne0$ affects the energy
density of neutrino via changing the distribution function function of
neutrino through nonzero chemical potential.  The lepton number density
and neutrino energy density are both written in terms of neutrino
asymmetry parameter $\xi$ as
\begin{equation}
   n_L=N_\nu\frac{\zeta(3)}{\pi^2}T_\nu^3\left[
   \frac{\xi}{\pi}+\left(\frac{\xi}{\pi}\right)^3\right] ,
\end{equation}
\begin{equation}
  \rho_\nu=N_\nu\frac{7\pi^2}{120}T_\nu^4
   \left[1+\frac{30}{7}\left(\frac{\xi}{\pi}\right)^2
    +\frac{15}{7}\left(\frac{\xi}{\pi}\right)^4\right].
\end{equation}
Keeping only leading terms in $\xi/\pi$, we can relate the neutrino
isocurvature perturbation $\mathcal{S}_\nu$ and the isocurvature
perturbation for lepton number density $\mathcal{S}_\mathrm{L}$ 
\footnote{
We simply assume there is no difference between perturbations in the
temperatures of neutrino and photon. This is because photon and neutrino
are thought to be coupled in the early universe at temperature $T\gtrsim
O(1)$ MeV and their temperature keep fluctuating in the same way after
the neutrino decoupling.
}

\begin{equation}
   \mathcal{S}_\nu=\frac{45}{7}\left(\frac{\xi}{\pi}\right)^2
    \mathcal{S}_\mathrm{L},
\end{equation}
which yields Eq.~(\ref{eq:cur_nu}).

More generally, the curvaton possibly decays before it completely
dominates the universe.  We therefore phenomenologically parametrize the
various perturbations by using $r_\mathrm{R}\equiv \rho_\sigma/\rho_T$,
the ratio of curvaton energy density just before its decay to the total
energy density just after the curvaton decay, and then initial
perturbations for the structure formation are written as
\footnote{%
Authors in~\cite{Ferrer:2004nv} used different parametrizations. Our
parametrizations $r_\mathrm{R}$ corresponds to $A_r$
in~\cite{Ferrer:2004nv} with taking $\lambda_m=\lambda_r=1$.
}
\begin{eqnarray}
   \zeta&=&\zeta_*+\frac{r_\mathrm{R}}{3}\mathcal{S}_\sigma\\[0.5em]
    \mathcal{S}_\mathrm{CDM}&=&(r_\mathrm{CDM}-r_\mathrm{R})
    \mathcal{S}_\sigma+f_\nu\mathcal{S}_\nu\\[0.5em]
    \mathcal{S}_b&=&
      (r_\mathrm{B}-r_\mathrm{R})
      \mathcal{S}_\sigma+f_\nu\mathcal{S}_\nu\\[0.5em]
    \mathcal{S}_\nu&=&
    \frac{45}{7}\left(\frac{\xi}{\pi}\right)^2
     (r_\mathrm{L}-r_\mathrm{R})\mathcal{S}_\sigma.
\end{eqnarray}
In the case the curvaton decays after it completely dominates the
universe and its energy turns into radiation nearly completely,
$r_\mathrm{R}=1$. 

Now we are prepared to obtain constraints on curvaton scenarios.  For
simplicity, we consider the case that the curvature perturbation
generated at the inflation epoch is negligible ($\zeta_*=0$) and the
curvature perturbation is created by the curvaton.  In that case the
isocurvature perturbation is completely correlated with the curvature
perturbation.  Then, we can represent the parameters $B_a^\prime$ 
in Eq.~(\ref{eq:Bprime}) as
\begin{eqnarray}
   B_\mathrm{CI}^\prime&=&
    -3\left(1-\frac{r_\mathrm{CDM}}{r_\mathrm{R}}\right)\\
   B_\mathrm{BI}^\prime&=&
    -3\left(1-\frac{r_\mathrm{B}}{r_\mathrm{R}}\right)\\
   B_\mathrm{NID}^\prime&=&
    -\frac{405}{28(1-f_\nu)}\left(\frac{\xi}{\pi}\right)^2
    \left(1-\frac{r_\mathrm{L}}{r_\mathrm{R}}\right)
\end{eqnarray}

Using the constraints on totally correlated isocurvature models without
tensor modes obtained in Section~\ref{subsec:totally_correlated} we
obtain the following limits on $B'$\footnote{
We here used the standard value for the neutrino fraction in the energy
density of the radiation, $f_\nu=0.40$. However if the large lepton
asymmetry exists the thermal history of the neutrino is modified so that
$f_\nu$ is changed and the structure formation is also affected.  We
refer readers to \cite{Lesgourgues:1999wu} for various effects of the
lepton asymmetry on the structure formation.  Here we assume that the
lepton number, if any, is sufficiently small and the thermal history of
the neutrino is not affected.  }:
\begin{eqnarray}
   -0.029\le&1- \frac{r_\mathrm{CDM}}{r_R}&\le 0.027 
    \label{eq:cur_rCDM} ,\\
   -0.134\le&1- \frac{r_\mathrm{B}}{r_R}&\le 0.133
    \label{eq:cur_rB}, \\
   -7.2\times10^{-3}\le& \left(\frac{\xi}{\pi}\right)^2
    \left(1-\frac{r_\mathrm{L}}{r_\mathrm{R}}\right)
   &\le 3.7\times10^{-3} .
    \label{eq:cur_rL}
\end{eqnarray}
These constraints on curvaton scenarios are slightly stringent compared
to those in~\cite{Gordon:2002gv}, and roughly same as those
in~\cite{Beltran:2004uv} and~\cite{Bean:2006qz}.

For $r_\mathrm{R}=1$, the constraints imply that both CDM and baryon
number should be created by or after the decay of the curvaton
($r_\mathrm{CDM}\simeq r_\mathrm{B}\simeq 1$).  On the other hand,
production of the lepton number is not constrained since no observation
at present indicates the presence of non zero lepton number in the
universe and $\xi$ is consistent to zero.  If we take a natural
assumption that the lepton number should be comparable to the baryon
number $n_\mathrm{L}/s \simeq n_\mathrm{B}/s \simeq 10^{-10}$ then the
neutrino asymmetry parameter $\xi$ should be of order $10^{-9}$. With
such a small value of $\xi$ the constraint Eq.~(\ref{eq:cur_rL}) then
leads to
\begin{equation}
   0\le \frac{r_L}{r_R}\le 10^6,
\end{equation}
and unless $r_\mathrm{R}\le 10^{-6}$ no restriction is assigned in
generation of lepton number.  Conversely, if nonzero fraction of the
neutrino isocurvature density fluctuation is favored by future
observations, the existence of large lepton number asymmetry may be
suggested.

We finally make a comment on  the case where both $\zeta_*$ and
$\mathcal{S}_\sigma$ contribute to the initial perturbations for the
structure formation.  In this case, the constraints are weakened 
by a factor $\sim \mathcal{S}_\sigma /(3\zeta_*/r_{\rm R}+
\mathcal{S}_\sigma)$.

\section{Conclusion}
\label{sec:conclusion}

We have presented constraints on isocurvature and tensor perturbations
from the combination of CMB and LSS data.  We have considered models
with one isocurvature mode (CI, NID or NIV) and tensor modes. As for
correlation of the isocurvature mode to the adiabatic mode, we have
investigated three models; uncorrelated, totally correlated and
generally correlated isocurvature models.

For totally correlated isocurvature models, the contribution of
isocurvature perturbation is severely limited $B_a\le 0.1\sim 0.2$.  For
uncorrelated and generally correlated isocurvature models we obtain
$B_a\le 0.3\sim 0.7$ and upper limits are a few times larger than those
for totally correlated models.  Compared with other recent constraints
on isocurvature models without tensor modes, our limits are roughly same
even if contribution of tensor modes is included.  

We have also obtained the upper limits on the tensor mode taking the
isocurvature mode into account.  The limits are strongly depends on the
isocurvature modes and its correlation included in the models.  For CI
and NID modes, constraints for uncorrelated isocurvature models are
similar to those for purely adiabatic models, but constraints weaken
when correlation with adiabatic modes are included.  For NIV modes,
constraints for both uncorrelated and totally correlated models are
similar as those for purely adiabatic models, but for generally
correlated model, constraints loosen significantly.

Finally we have found no significant improvement of $\chi^2$ for models
with isocurvature and tensor mode.  Thus we conclude the initial
conditions of the structure formation are still consistent with
completely adiabatic ones.

We have also applied the obtained constraints to some specific models
which leads to the isocurvature perturbations, the axion isocurvature
perturbation model and the curvaton scenario.  Since the axion decay
constant is bounded around $10^{11}$ GeV, the scale of inflation
$H_\mathrm{inf}$ which determines the amplitude of the axion
fluctuation is constrained to be below $10^7$ GeV.  Thus, very low scale
inflation is required.  As for the curvaton scenario, when the curvaton
dominated the universe before its decay, we have shown that CDM and
baryon number observed in the present universe should be created by or
after the decay of the curvaton, otherwise too large isocurvature
fluctuation is produced. However the generation of lepton number is not
constrained by current cosmological observations.

\bigskip 
\noindent
{\bf Acknowledgment:} 
We  would like to thank Kazuhide Ichikawa for useful comments and discussions.
This work was supported in part by the Grant-in-Aid for Scientific Research
from the Ministry of Education, Science, Sports, and Culture of Japan,
No. 18540254 and No 14102004 (M.K.).  This work was also supported
in part by JSPS-AF Japan-Finland Bilateral Core Program (M.K.)


\begin{thebibliography}{100}

\bibitem{Spergel:2006hy}
  D.~N.~Spergel {\it et al.}  [WMAP Collaboration],
  arXiv:astro-ph/0603449.

\bibitem{Tegmark:2006az}
  M.~Tegmark {\it et al.},
  Phys.\ Rev.\  D {\bf 74}, 123507 (2006)

\bibitem{Lyth:2001nq}
  D.~H.~Lyth and D.~Wands,
  Phys.\ Lett.\  B {\bf 524}, 5 (2002)
  
\bibitem{Moroi:2001ct}
  T.~Moroi and T.~Takahashi,
  Phys.\ Lett.\  B {\bf 522}, 215 (2001)
  [Erratum-ibid.\  B {\bf 539}, 303 (2002)]

\bibitem{Pierpaoli:1999zj}
  E.~Pierpaoli, J.~Garcia-Bellido and S.~Borgani,
  JHEP {\bf 9910}, 015 (1999)

\bibitem{Enqvist:2000hp}
  K.~Enqvist, H.~Kurki-Suonio and J.~Valiviita,
  Phys.\ Rev.\  D {\bf 62}, 103003 (2000)
  
\bibitem{Trotta:2001yw}
  R.~Trotta, A.~Riazuelo and R.~Durrer,
  Phys.\ Rev.\ Lett.\  {\bf 87}, 231301 (2001)

\bibitem{Trotta:2002iz}
  R.~Trotta, A.~Riazuelo and R.~Durrer,
  Phys.\ Rev.\  D {\bf 67}, 063520 (2003)

\bibitem{Valiviita:2003ty}
  J.~Valiviita and V.~Muhonen,
  Phys.\ Rev.\ Lett.\  {\bf 91}, 131302 (2003)

\bibitem{Crotty:2003rz}
  P.~Crotty, J.~Garcia-Bellido, J.~Lesgourgues and A.~Riazuelo,
  Phys.\ Rev.\ Lett.\  {\bf 91}, 171301 (2003)

\bibitem{Moodley:2004nz}
  K.~Moodley, M.~Bucher, J.~Dunkley, P.~G.~Ferreira and C.~Skordis,
  Phys.\ Rev.\  D {\bf 70}, 103520 (2004)

\bibitem{Beltran:2004uv}
  M.~Beltran, J.~Garcia-Bellido, J.~Lesgourgues and A.~Riazuelo,
  Phys.\ Rev.\  D {\bf 70}, 103530 (2004)

\bibitem{Bean:2006qz}
  R.~Bean, J.~Dunkley and E.~Pierpaoli,
  Phys.\ Rev.\  D {\bf 74}, 063503 (2006)

\bibitem{Trotta:2006ww}
  R.~Trotta,
  Mon.\ Not.\ Roy.\ Astron.\ Soc.\ Lett.\  {\bf 375}, L26 (2007)

\bibitem{Keskitalo:2006qv}
  R.~Keskitalo, H.~Kurki-Suonio, V.~Muhonen and J.~Valiviita,
  arXiv:astro-ph/0611917.

\bibitem{Bartolo:2001rt}
  N.~Bartolo, S.~Matarrese and A.~Riotto,
  Phys.\ Rev.\  D {\bf 64}, 123504 (2001)

\bibitem{Byrnes:2006fr}
  C.~T.~Byrnes and D.~Wands,
  Phys.\ Rev.\  D {\bf 74}, 043529 (2006)

\bibitem{Bucher:1999re}
  M.~Bucher, K.~Moodley and N.~Turok,
  Phys.\ Rev.\  D {\bf 62}, 083508 (2000)

\bibitem{Hinshaw:2006ia}
  G.~Hinshaw {\it et al.}  [WMAP Collaboration],
  arXiv:astro-ph/0603451.

\bibitem{Page:2006hz}
  L.~Page {\it et al.}  [WMAP Collaboration],
  arXiv:astro-ph/0603450.

\bibitem{Cole:2005sx}
  S.~Cole {\it et al.}  [The 2dFGRS Collaboration],
  Mon.\ Not.\ Roy.\ Astron.\ Soc.\  {\bf 362} (2005) 505

\bibitem{Lewis:1999bs}
  A.~Lewis, A.~Challinor and A.~Lasenby,
  Astrophys.\ J.\  {\bf 538}, 473 (2000)

\bibitem{Lewis:2002ah}
  A.~Lewis and S.~Bridle,
  Phys.\ Rev.\  D {\bf 66}, 103511 (2002)

\bibitem{Gelman:1992}
 A.~Gelman and D.~Rubin,
  Statistical\ Science\ {\bf 7}, 457 (1992)

\bibitem{Bucher:2000cd}
  M.~Bucher, K.~Moodley and N.~Turok,
  arXiv:astro-ph/0011025.

\bibitem{Peccei:1977hh}
  R.~D.~Peccei and H.~R.~Quinn,
  Phys.\ Rev.\ Lett.\  {\bf 38}, 1440 (1977).

\bibitem{Weinberg:1977ma}
  S.~Weinberg,
  Phys.\ Rev.\ Lett.\  {\bf 40}, 223 (1978).

\bibitem{Wilczek:1977pj}
  F.~Wilczek,
  Phys.\ Rev.\ Lett.\  {\bf 40}, 279 (1978).

\bibitem{Turner:1989vc}
  M.~S.~Turner,
  Phys.\ Rept.\  {\bf 197}, 67 (1990).

\bibitem{Raffelt:1990yz}
  G.~G.~Raffelt,
  Phys.\ Rept.\  {\bf 198} (1990) 1.

\bibitem{Seckel:1985tj}
  D.~Seckel and M.~S.~Turner,
  Phys.\ Rev.\  D {\bf 32}, 3178 (1985).

\bibitem{Turner:1990uz}
  M.~S.~Turner and F.~Wilczek,
  Phys.\ Rev.\ Lett.\  {\bf 66}, 5 (1991).

\bibitem{Steinhardt:1983ia}
  P.~J.~Steinhardt and M.~S.~Turner,
  Phys.\ Lett.\  B {\bf 129}, 51 (1983).


\bibitem{Kawasaki:1995ta}
  M.~Kawasaki, N.~Sugiyama and T.~Yanagida,
  Phys.\ Rev.\  D {\bf 54}, 2442 (1996)
  
\bibitem{Kawasaki:1997ct}
  M.~Kawasaki and T.~Yanagida,
  Prog.\ Theor.\ Phys.\  {\bf 97}, 809 (1997)

\bibitem{Kasuya:1997td}
  S.~Kasuya, M.~Kawasaki and T.~Yanagida,
  Phys.\ Lett.\  B {\bf 415}, 117 (1997)
  
\bibitem{Burns:1997ue}
  S.~D.~Burns,
  arXiv:astro-ph/9711303.

\bibitem{Kanazawa:1998pa}
  T.~Kanazawa, M.~Kawasaki, N.~Sugiyama and T.~Yanagida,
  Prog.\ Theor.\ Phys.\  {\bf 100}, 1055 (1998)

\bibitem{Lyth:1989pb}
  D.~H.~Lyth,
  Phys.\ Lett.\  B {\bf 236}, 408 (1990).

\bibitem{Beltran:2006sq}
  M.~Beltran, J.~Garcia-Bellido and J.~Lesgourgues,
  arXiv:hep-ph/0606107.

\bibitem{Gross:1980br}
  D.~J.~Gross, R.~D.~Pisarski and L.~G.~Yaffe,
  Rev.\ Mod.\ Phys.\  {\bf 53}, 43 (1981).

\bibitem{Weinberg:1996kr}
  S.~Weinberg,
  ``The quantum theory of fields. Vol. 2: Modern applications,''
{\it  Cambridge, UK: Univ. Pr. (1996) 489 p}

\bibitem{Yao:2006px}
  W.~M.~Yao {\it et al.}  [Particle Data Group],
  J.\ Phys.\ G {\bf 33}, 1 (2006).

\bibitem{Serpico:2005bc}
  P.~D.~Serpico and G.~G.~Raffelt,
  Phys.\ Rev.\  D {\bf 71}, 127301 (2005)

\bibitem{Olive:2004kq}
  K.~A.~Olive and E.~D.~Skillman,
  Astrophys.\ J.\  {\bf 617}, 29 (2004)

\bibitem{Lesgourgues:1999wu}
  J.~Lesgourgues and S.~Pastor,
  Phys.\ Rev.\  D {\bf 60}, 103521 (1999)

\bibitem{Lyth:2002my}
  D.~H.~Lyth, C.~Ungarelli and D.~Wands,
  Phys.\ Rev.\  D {\bf 67}, 023503 (2003)

\bibitem{Ferrer:2004nv}
  F.~Ferrer, S.~Rasanen and J.~Valiviita,
  JCAP {\bf 0410}, 010 (2004)

\bibitem{Gordon:2002gv}
  C.~Gordon and A.~Lewis,
  Phys.\ Rev.\  D {\bf 67}, 123513 (2003)





\end{thebibliography}
\end{document}